\documentclass[letterpaper, 10 pt, conference]{ieeeconf}  

\IEEEoverridecommandlockouts                              

\usepackage{graphics} 
\usepackage{epsfig} 
\usepackage{mathptmx} 
\usepackage{times} 
\usepackage{amsmath} 
\usepackage{amssymb}  
\usepackage{cite}
\usepackage{xcolor}

\newtheorem{remark}{Remark}[section]

\title{\LARGE \bf
Physics--Guided Neural Networks for Inversion--based \\ Feedforward Control applied to Linear Motors
}

\author{Max Bolderman, Mircea Lazar and Hans Butler
\thanks{\begin{flushright}
\begin{minipage}[r]{0.04\textwidth}
    \includegraphics[width=0.9\linewidth]{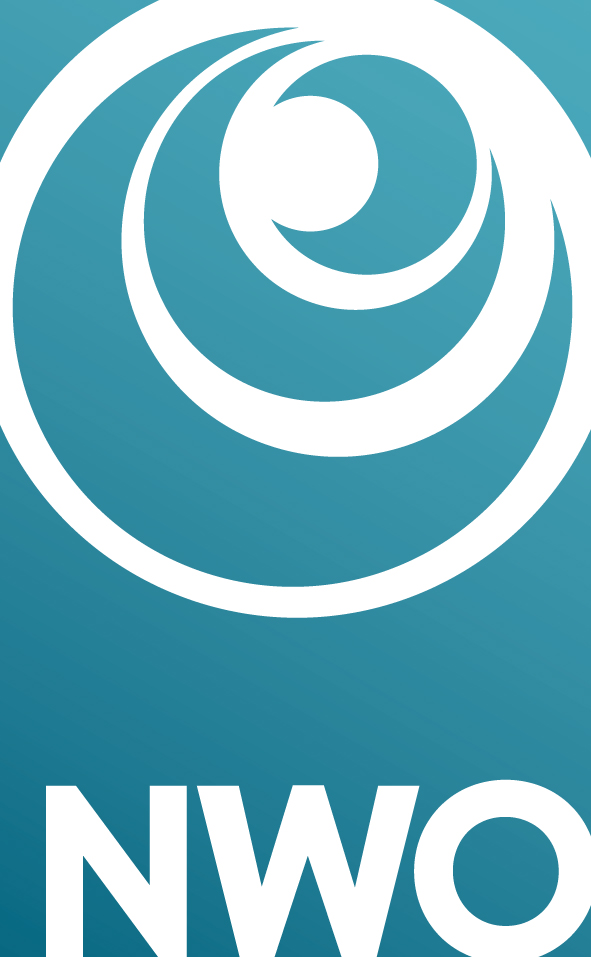}
\end{minipage}
\end{flushright}
\vspace{-1.05cm}
\begin{minipage}[l]{0.43\textwidth}
\hspace{0.5em} *This work is part of the research programme 9654 with project number 17973, which is (partly) financed by the Dutch Research Council (NWO). 
\end{minipage}
}
\thanks{All authors are with the Department of Electrical Engineering, Eindhoven University of Technology, The Netherlands, E-mails: m.bolderman@tue.nl, m.lazar@tue.nl and h.butler@tue.nl.}%
}

\begin{document}

\maketitle
\thispagestyle{empty}
\pagestyle{empty}


\begin{abstract}
Ever--increasing throughput specifications in semiconductor manufacturing require operating high--precision mechatronics, such as linear motors, at higher accelerations. In turn this creates higher nonlinear parasitic forces that cannot be handled by industrial feedforward controllers. Motivated by this problem, in this paper we develop a general framework for inversion--based feedforward controller design using physics--guided neural networks (PGNNs). In contrast with black--box neural networks, the developed PGNNs embed prior physical knowledge in the input and hidden layers, which results in improved training convergence and learning of underlying physical laws. The PGNN inversion--based feedforward control framework is validated in simulation on an industrial linear motor, for which it achieves a mean average tracking error twenty times smaller than mass--acceleration feedforward in simulation. 
\end{abstract}


\section{Introduction}
\label{sec:Introduction}
In semiconductor manufacturing there is a constant drive to maximize throughput, which translates into improving accuracy of high--precision mechatronics, such as industrial coreless linear motors (CLMs) \cite{Rohrig2005, Nguyen2018}. This calls for operating CLMs at higher velocities and accelerations, while further reducing the position error during motion. However, higher accelerations result in higher parasitic nonlinear friction and electromagnetic forces. 
Typically, feedforward control is responsible for compensating these forces and achieving high accuracy of the position control loop \cite{Steinbuch2000}, while feedback control provides closed--loop stability and corrects remaining small tracking errors. Therefore, there is an interest in developing advanced feedforward controllers that can effectively compensate nonlinear parasitic forces.  



Several methods for feedforward controller design have been proposed in literature. A well--known type is inversion--based feedforward control, which uses a parametric model of the inverse system dynamics. The standard industrial mass--acceleration feedforward controllers are of this type, and they can be extended to compensate linear friction forces \cite{Nguyen2018}. In general, inverse models can be obtained by inversion of the forward system dynamics, e.g., \cite{Butterworth2012}, or by directly approximating inverse dynamics, e.g., \cite{Sorensen1999, Blanken2020}. Performance of inversion--based feedforward controllers is limited by the accuracy of the model inversion and quality of the inverse model \cite{Boerlage2003, Zundert2018}. Black--box feedforward neural networks (NNs) have been originally employed in inversion--based feedforward control in \cite{Sorensen1999} due to their universal approximating capabilities \cite{Hornik1989}, with a validation on a chemical process. Recently, \cite{Gilra2018} employed more advanced types of NNs (spiking and differential NNs) for nonlinear model inversion--based control in a robotic arm. 

A different popular approach to feedforward controller design is iterative learning control (ILC), which achieves superior performance by iteratively learning a control input for one repeating task \cite{ILC}. Extrapolation of the learned control input to a different task however, yields a deterioration in performance \cite{Heertjes2009, Bolder2014}. The use of interpolation and non--causal rational basis functions to parameterize feedforward ILCs has resulted in improved ILCs, which are robust against varying references, see \cite{Blanken2020b} and the references therein.
Utilization of NNs in combination with ILC for feedforward motion control was originally suggested in \cite{Steinbuch2000} and has been recently exploited in \cite{Patan2017} to extend applicability of ILC to nonlinear systems. 

Despite the above--mentioned promising results that exploit black--box NNs in inversion--based or ILC feedfoward control, training NNs to approximate general system dynamics remains troublesome even when using techniques such as cross--validation or Bayesian regularization. This hinders the usage of NN--based controllers in industrial high--precision mechatronics, where failure to correctly learn underlying physical laws can result in unsafe control inputs.



Motivated by this bottleneck, in this paper we develop a general framework for inversion-–based  feedforward  controller  design  using  physics–-guided  neural  networks  (PGNNs).  In  contrast  with  black–-box NNs,  the  developed  PGNNs  allow embedding  prior physical knowledge  in  the  input  and  hidden  layers, which  results in a NN with a hybrid structure. For example, the hidden layer may consist of two parts, a physics--guided part that inherits the structure of a known physics--based model, and a black--box part, which is useful for learning any unknown parasitic forces. The resulting PGNN with a hybrid, parallel structure, is then trained as a single NN, with both parts of the hidden layer contributing to the NN output. 
The developed PGNN inversion--based feedforward control framework is validated in simulation for position control of a industrial linear motor with nonlinear bearing friction, where it outperforms alternative inversion--based feedforward controllers.


\begin{remark} An alternative type of physics--guided \emph{black--box} NNs was proposed in \cite{Karpatne2017} and applied to a lake temperature estimation problem. Therein, the physical guidance consists of using a physics--based loss function (cost to be minimized during training), which penalizes the deviation of NN outputs from compliance with an available physics--based model. This approach has also been termed as physics--informed neural networks (PINNs) in \cite{Stiasny2020} (see also the references therein), where it was applied to identification of nonlinear power systems dynamics. 
Recently, a physics--guided architecture for NNs was also proposed in \cite{Karpatne2019} and applied to a lake temperature estimation problem. Therein, measured physical variables are assigned to certain neuron outputs and used in the physics--based loss function. Additionally, neural connections are designed to have these physical variables follow basic physics--based relationships. 
The PGNN type of NN developed in this paper differs by using a physics--guided input transformation in combination with a hybrid, parallel structure for the hidden layer consisting of a black--box part and a physics--guided part. Another important difference is that the resulting PGNN is trained as a standard NN, without using a physics--based loss function. 
\end{remark}

\begin{remark} An inversion--based feedforward controller for compensating friction forces in an industrial CLM was recently developed in \cite{Yuen2019} using a \emph{linear} NN representation, i.e., a simple input--output graph structure without a hidden layer that is equivalent with an autoregressive exogonenous linear model \cite{Hof2020}. Additionally, in \cite{Yuen2019b} an extension was proposed that enables the compensation of the nonlinear Coulomb friction using a similar \emph{linear} NN. Training is based on solving a least square optimization problem, but the approach therein is not able to compensate for unmodeled parasitic forces, as illustrated in Section~\ref{sec:4}.
\end{remark}




\section{Preliminaries and Problem Formulation}
\label{sec:Preliminaries}
\begin{figure}
	\centering
	\includegraphics[width=1\linewidth]{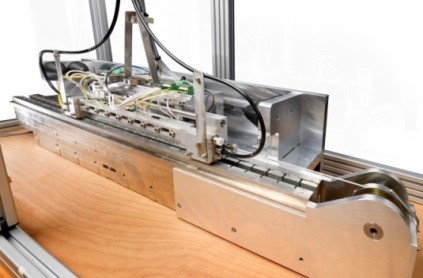}
	\caption{Experimental CLM setup.}
	\label{fig:CLM}
\end{figure}

\subsection{CLM Setup and Dynamics}
The experimental CLM setup is displayed in Fig. \ref{fig:CLM}, and is commonly modeled as a mechanical system preceded by an electromagnetic nonlinearity \cite{Nguyen2018}. Typically, a commutation algorithm is used to invert the electromagnetics, such that the position controller design can be done for the mechanical dynamics independently. 
The power electronics, responsible for controlling the currents in the electromagnetic subsystem, are in general significantly faster than their mechanical counterpart, and therefore neglected for controller design purposes \cite{Rohrig2005}.

\begin{figure}
	\centering
	\includegraphics[width=1\linewidth]{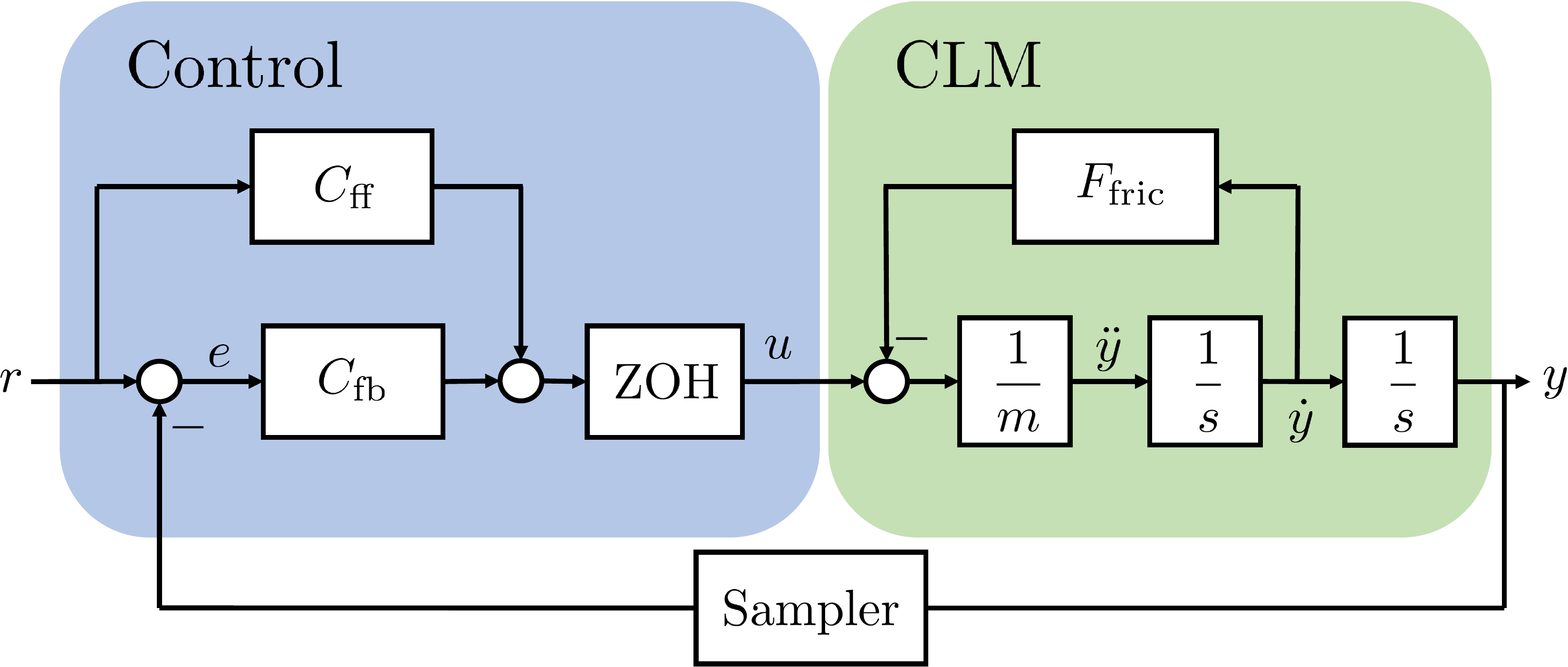}
	\caption{Control schematics for the mechanical dynamics of the CLM.}
	\label{fig:CLMOverview}
\end{figure}

The control schematics for the mechanical dynamics of a CLM are presented in Fig. \ref{fig:CLMOverview}. Here, $r$ is the reference position, $y$ the actual position, and $e := r-y$ the tracking error. The feedforward ($C_{\textup{ff}}$) and feedback ($C_{\textup{fb}}$) controllers compute the force input $u$ and they are implemented in discrete time as indicated by the ZOH and Sampler blocks.
Typically, reference values are known beforehand, such that these can be used by the feedforward controller.
The dynamics of the CLM are modeled as
\begin{equation}
    \label{eq:CLM}
	m \ddot{y} = u - F_{\textup{fric}},
\end{equation}
with $m$ the moving mass, and $F_{\textup{fric}}$ the nonlinear mechanical friction. The dot above a variable indicates a differentiation with respect to time, i.e., $\dot{y} = \frac{d y}{dt}$.
In \cite{Nguyen2018}, the friction model
\begin{align}
\begin{split}
\label{eq:FrictionModel}
	F_{\textup{fric}}(\dot{y}, y) = &f_v \dot{y} + f_c \textup{sign}(\dot{y}) + (f_s -f_c) \textup{sign}(\dot{y}) e^{- \left( \frac{\dot{y}}{v_s} \right)^2} \\
	& + c_1 \sin(\omega y ),
\end{split}
\end{align}
is identified using data from the CLM depicted in Fig.~\ref{fig:CLM}. Here, $f_v$, $f_c$, $f_s$, and $c_1$ are the viscous, Coulomb, Stribeck and sinusoidal friction coefficients, respectively. Additionally, $v_s$ denotes the Stribeck velocity and $\omega$ is the period of the position dependent sinusoidal friction. Parameter values are summarized in Table~\ref{tab:Parameters}.
\begin{table}
	\caption{Parameters identified from the CLM measured data.}
	\begin{center}
		\begin{tabular}{ccc}
			\hline
			{\textbf{Parameter}} & {\textbf{Value}} & {\textbf{Unit}} \\
			\hline \hline
			$f_c$& 8.72 & $N$   \\
			$f_v$& 41.22 & $\frac{N}{m/s}$\\
			$f_s$& 14.63 & $N$\\
			$v_s$& $1.23\cdot 10^{-3}$ & $m/s$\\
			$c_1$& -1.44 & $N$\\
			$\omega$& 2.21 & $rad/m$\\
			$m$&19.96&$kg$\\
			\hline
		\end{tabular}
	\end{center}
	\label{tab:Parameters}
\end{table}
Although the actual friction force is setup specific, \eqref{eq:FrictionModel} represents a general static friction model.

Due to the possible drift in position, CLMs cannot be operated in open--loop. To this end, the low--pass filtered PID feedback controller
\begin{equation}
\label{eq:Feedback}
	C_{\textup{fb}}(s) = 6600 \left( \frac{s+ \frac{5}{6} 2 \pi}{s} + \frac{s+ \frac{5}{3} 2 \pi}{s+ 30 \pi} \right)\frac{1}{s+ 60 \pi}
\end{equation}
was tuned based on a simplified friction model $F_{\textup{fric}}(\dot{y}) = f_v \dot{y}$, using standard loop--shaping techniques \cite{Steinbuch2000}. The controller was then discretized using sampling time $T_s = 10^{-4}$ $s$.

\subsection{Benchmark Feedforward Control Strategies}
\label{sec:Benchmarks}
Consider the nonlinear discrete--time inverse dynamics 
\begin{equation}
    \label{eq:InverseDynamics}
    u(t) = f(y(t+n_a), \hdots, y(t-n_b), u(t-1), \hdots, u(t-n_c), \hdots),
\end{equation}
with time instant $t = kT_s$, where $k= 0,1,\hdots$. More dependencies can be added as indicated with the dots at the end. The general discrete--time inversion--based feedforward controller is then obtained by using $r(t)$, whose future values are known, instead of $y(t)$ in \eqref{eq:InverseDynamics}, which gives
\begin{equation}
    \label{eq:GeneralFeedforward}
    u(t) = f(r(t+n_a), \hdots, r(t-n_b), u(t-1), \hdots, u(t-n_c), \hdots). 
\end{equation}
Previous values of the feedforward input are used in \eqref{eq:GeneralFeedforward}, such that the feedforward is not part of the feedback loop, see Fig.~\ref{fig:CLMOverview}. 
In general, \eqref{eq:InverseDynamics} is used offline to identify the inverse dynamics, after which \eqref{eq:GeneralFeedforward} gives the inversion--based feedforward controller.

To facilitate the definition of benchmark feedforward controllers, we specify some structure in \eqref{eq:GeneralFeedforward}. 
Due to the small sampling time $T_s = 10^{-4}$ $s$, we assume that $F_{\textup{fric}}$ defined in \eqref{eq:FrictionModel} is constant in between two consecutive samples.
This justifies a ZOH discretization of \eqref{eq:CLM}, such that
\begin{align}
\begin{split}
\label{eq:FCStep2}
	y(t) & = \frac{T_s^2}{2m} \frac{q^{-1} + q^{-2} }{(1-q^{-1})^2} (u(t) - F_{\textup{fric}}(y(t), \dot{y}(t)).
\end{split}
\end{align}
where $q^{-1}$ is the backwards shift operator and, with a slight abuse of notation, $\dot{y}(t)$ stands for the discretization operator $\frac{q-q^{-1}}{2T_s} y(t)$. Multiplying both sides of \eqref{eq:FCStep2} with $q$, rewriting and setting $r(t) = y(t)$, gives the general CLM feedforward controller
\begin{align}
\begin{split}
    \label{eq:CLMFeedforward}
    u(t) =&  -q^{-1}u(t) + \frac{2m}{T_s^2} q(1-q^{-1})^2 r(t) + \\
    & +(1+q^{-1}) F_{\textup{fric}}(r(t), \dot{r}(t)).
\end{split}
\end{align}
The three benchmark feedforwards are next introduced by considering special cases of \eqref{eq:CLMFeedforward} and using estimated parameter values (i.e., $\hat{\theta}$ is an estimate of $\theta$).  

The first benchmark, mass--acceleration feedforward, assumes zero friction and compensates only for the inertial forces. Substituting $F_{\textup{fric}} = 0$ in \eqref{eq:CLMFeedforward} and using an estimated mass $\hat{m}$, gives the feedforward controller
\begin{equation}
\label{eq:MassFeedforward}
    u(t) = -q^{-1} u(t) + \frac{2 \hat{m}}{T_s^2} q( 1 - q^{-1})^2 r(t).
\end{equation}

The second benchmark is the friction compensator proposed in \cite{Yuen2019b}, which assumes a combination of viscous and Coulomb friction. Substituting $F_{\textup{fric}} (\dot{r}(t)) = \hat{f}_v \dot{r}(t) + \hat{f}_c \textup{sign}(\dot{r}(t))$ in \eqref{eq:CLMFeedforward}, and using an estimated mass $\hat{m}$, gives the feedforward controller 
\begin{align}
\begin{split}
\label{eq:FrictionCompensatorFeedforward}
	u(t) =& - q^{-1} u(t) + \frac{2 \hat{m}}{T_s^2} \left( 1 - q^{-1} \right)^2 q r(t) \\
	& + (1+q^{-1}) \left( \hat{f}_v \dot{r}(t) + \hat{f}_c \textup{sign} (\dot{r}(t)) \right).
\end{split}
\end{align}
The main contribution of \cite{Yuen2019b} is the use of the nonlinear transformation $\textup{sign}(\dot{y}(t))$ within the identification procedure of $m$, $f_v$, and $f_c$. This enables computation of optimal parameter values from available input--output data by convex optimization. 

\begin{figure}
	\centering
	\includegraphics[width=0.9\linewidth]{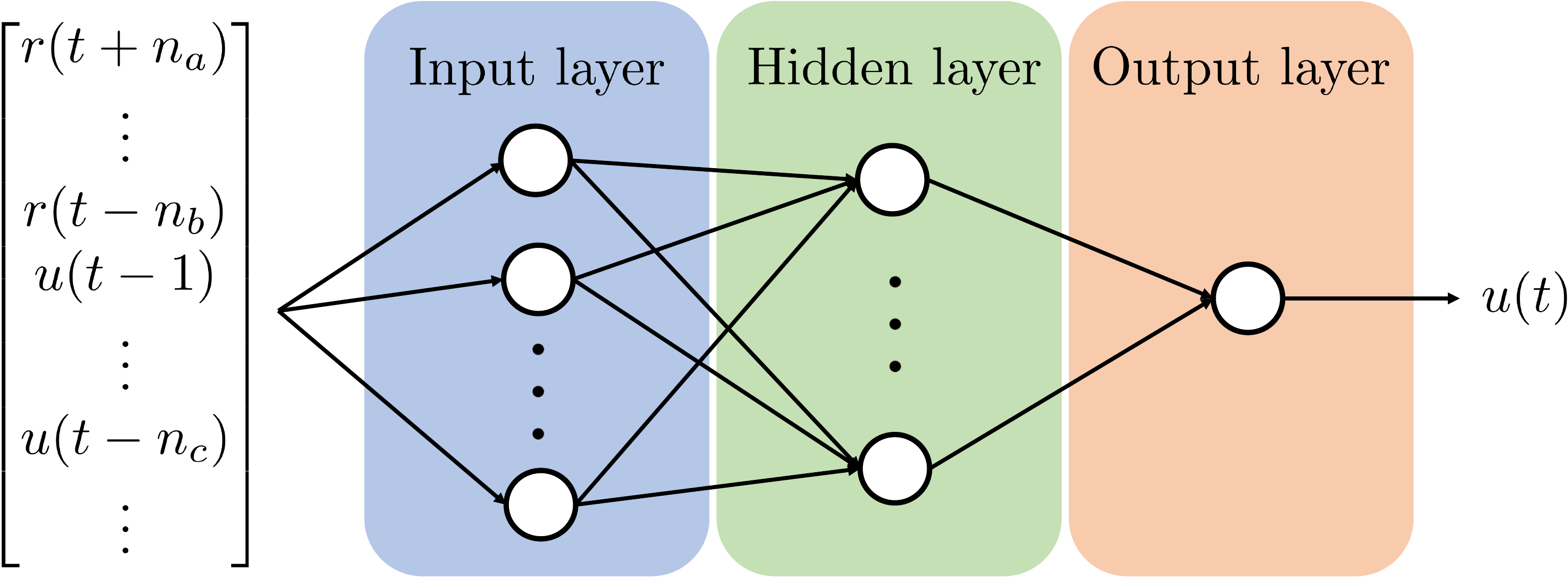}
	\caption{Black--box NNARX structure with a single hidden layer.}
	\label{fig:NNARX}
\end{figure}

The third benchmark is the inversion--based NN feedforward controller proposed in \cite{Sorensen1999}. A black--box multi--layer perceptron (MLP), see Fig. \ref{fig:NNARX}, is used to identify the inverse system dynamics \eqref{eq:FCStep2}. The method is denoted as NNARX referring to a black--box \emph{nonlinear} neural network model with autoregressive exogeneous (ARX) structure. A rough order estimation using \eqref{eq:FrictionCompensatorFeedforward} gives the NNARX feedforward controller
\begin{equation}
\label{eq:NNARXFeedforward}
    u(t) = f_{\textup{MLP}} \left( r(t+1), r(t), r(t-1), r(t-2), u(t-1) \right),
\end{equation}
where $f_{\textup{MLP}}(\cdot)$ is obtained by recursively computing the output $x_i$ of layer $i$ using
\begin{equation}
    \label{eq:OutputMLP}
    x_i = \alpha_i \left( W_{i,i-1} x_{i-1} + B_i \right),
\end{equation}
with $\alpha_i(\cdot)$, $W_{i,i-1}$, and $B_i$ the activation function vector, weight matrix, and bias vector of layer $i$, respectively. 

\begin{figure}
	\centering
	\includegraphics[width=0.8\linewidth]{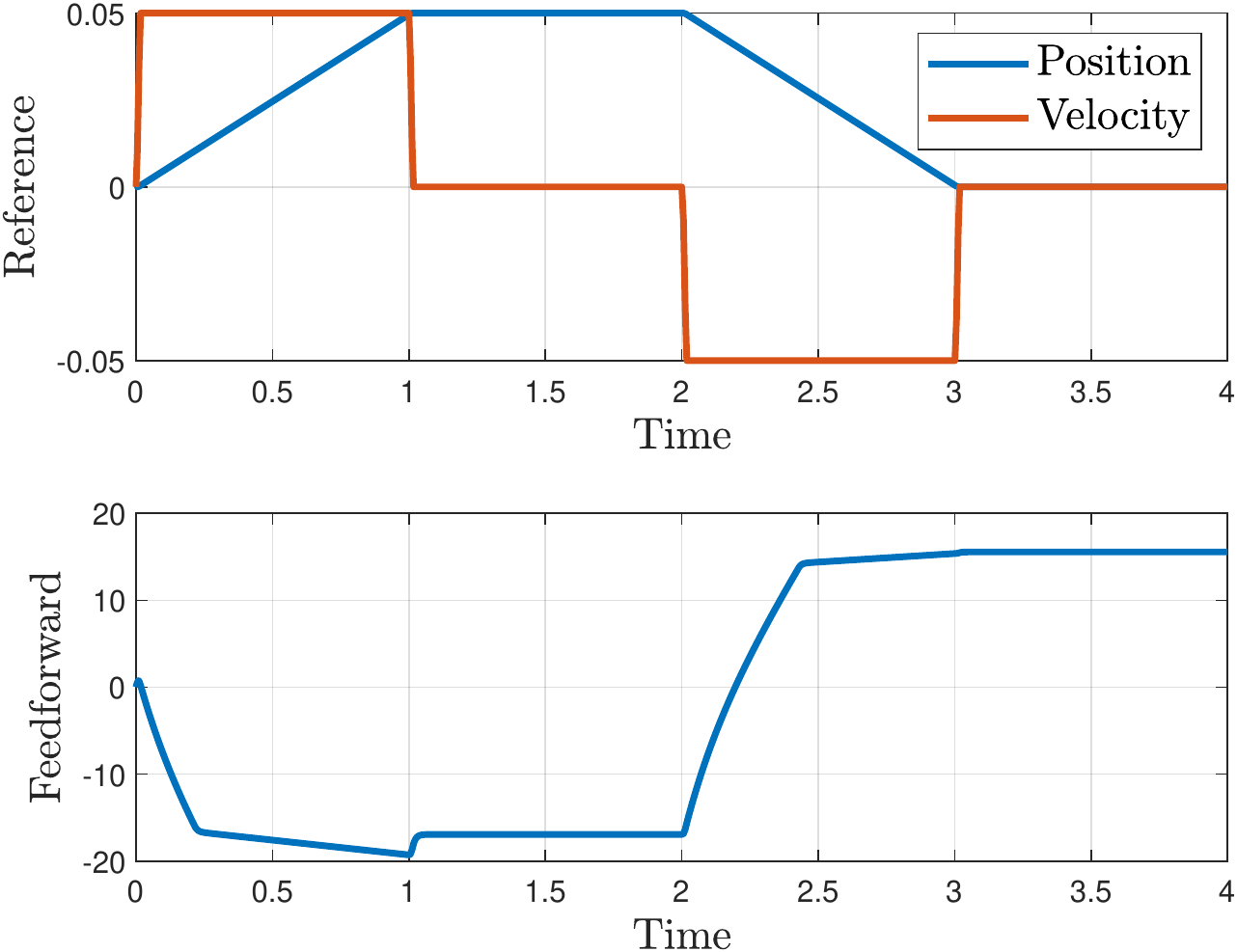}
	\caption{Generated feedforward input using a four--neuron ReLU NNARX.}
	\label{fig:NNARXFlawed}
\end{figure}

Following common practice, we choose linear activation functions $\alpha(x) = x$ for the input and output layer neurons, ReLU $\alpha(x) = \max(0, x)$ or tan-sigmoid $\alpha(x) = \frac{1}{1+e^{-2x}}-1$ activation functions for the hidden layer neurons, and we normalize the inputs and outputs within the domain $\left[ -1, 1 \right]$ to circumvent the vanishing gradient problem encountered when using sigmoid activation functions. 
Training the NN, i.e., the process of tuning the weights and biases, is performed with the Levenberg-Marquardt backpropagation algorithm and quadratic cost function
\begin{equation}
    \label{eq:CostFunction}
    V(w, Z^N):= \frac{1}{N} \sum^{N} \varepsilon(t)^2,
\end{equation}
with network weights and biases $w$, input--output data set of $N$ samples $Z^N= \left\{ u(0), y(0), \hdots , u(N-1), y(N-1) \right\}$, and estimation error $\varepsilon(t) = u(t) - \hat{u}(t)$ with $\hat{u}(t)$ the NN output on the training data. Notice that here $u$ denotes the applied force and $y$ the measured position. The identification criterion is given as
\begin{equation}
    \label{eq:IdentificationCriterium}
    \hat{w} = \textup{arg} \min_{w} V(w, Z^N).
\end{equation}
The non--convexity of this optimization makes it impossible to guarantee convergence to a global minimum. In practice, this might cause the NNARX to fail to learn general structures, rendering it practically infeasible as a feedforward controller. 
An example is presented in Fig.~\ref{fig:NNARXFlawed}, which shows the feedforward signal generated by a NNARX with four ReLU neurons in the hidden layer. Notice that from \eqref{eq:CLM}, we obtain the continuous--time optimal feedforward by substituting $y = r$, such that
\begin{equation}
\label{eq:IdealFF}
	u = F_{\textup{fric}} (r, \dot{r} ) + m \ddot{r}.
\end{equation}
A general observation from \eqref{eq:IdealFF} is that the feedforward input and reference velocity should have the same sign when the reference velocity is constant, i.e., $\ddot{r} = 0$ (note $f_c > \mid c_1 \mid$, Table~\ref{tab:Parameters}). 
Fig. \ref{fig:NNARXFlawed} clearly shows that the four--neuron ReLU NNARX failed to learn this basic relation, see, e.g., the feedforward input during $t \in \left(0, 1 \right)$. 
This exposes fragility of the black--box NNARX feedforward controllers.

\subsection{Problem Formulation}
The mass--acceleration feedforward and friction compensator described in the previous section are subject to limited performance due to the chosen model structure. The NNARX theoretically could overcome these limitations, but due to convergence to local minima in training it may fail to learn underlying physical laws, as shown in Fig.~\ref{fig:NNARXFlawed}.

Motivated by the above assessment, the problem considered in this paper is to improve the design and performance of inversion--based neural feedforward controllers by embedding physical knowledge within neural networks. To this end, we develop physics--guided neural networks that exploit the known physical dynamics of CLMs. The resulting PGNN feedforward controllers are validated on a industrial linear motor with the aim to improve tracking performance to $10$~$\mu m$ in the presence of the nonlinear friction \eqref{eq:FrictionModel}.




\section{Physics--Guided Neural Feedforward Controller Design}
\label{sec:MainResults}
This section describes the training data generation, and introduces and evaluates the newly proposed PGNN--I and PGNN--II feedforward controllers.

\subsection{Training Data Generation}
\label{sec:TrainingData}
For consistent closed--loop linear system identification, it suffices to have the model set capture the actual dynamics, and have a persistently exciting reference signal of an order at least equal to the number of coefficients to be estimated \cite{Hof2020}. 
For nonlinear system identification, \cite{Schoukens2019} states that the system should be excited close to its operating range. 

In order to identify the dynamics of the considered CLM we perform an experiment with the following settings:
\begin{enumerate}
    \item Dithering the input $u$, see Fig. \ref{fig:CLMOverview}, using a zero--mean white noise process with variance $(80)^2$ $N^2$ at a frequency of $100$ $Hz$. Values are kept constant inbetween two consecutive samples. 
    \item A third--order reference trajectory with desired position $r = \left\{0, 0.05 \right\}$ $m$, maximum velocity $v_d = 0.05$ $\frac{m}{s}$, maximum acceleration $a_d = 4$ $\frac{m}{s^2}$, and maximum jerk $j_d = 1000$ $\frac{m}{s^3}$, operating at a constant velocity $50$\% of the time, see Fig. \ref{fig:NNARXFlawed}.
\end{enumerate}
Dithering the CLM input directly prevents the signal from being filtered by the feedback controller and the lower frequency causes the CLM to explore more frequencies rather than sticking close to the reference trajectory.

Data is generated by sampling the input $u(t)$ and output $y(t)$ during a closed--loop simulation of four back--and--forth motions, with $T_s = 10^{-4}$~$s$. The data set is divided into $70$\% training, $15$\% validation, and $15$\% testing data. Cross-validation is performed using the validation set and training convergence is evaluated on the testing set.

\subsection{PGNN-I: Physics--Guided Input Design}
\begin{figure}
	\centering
	\includegraphics[width=1\linewidth]{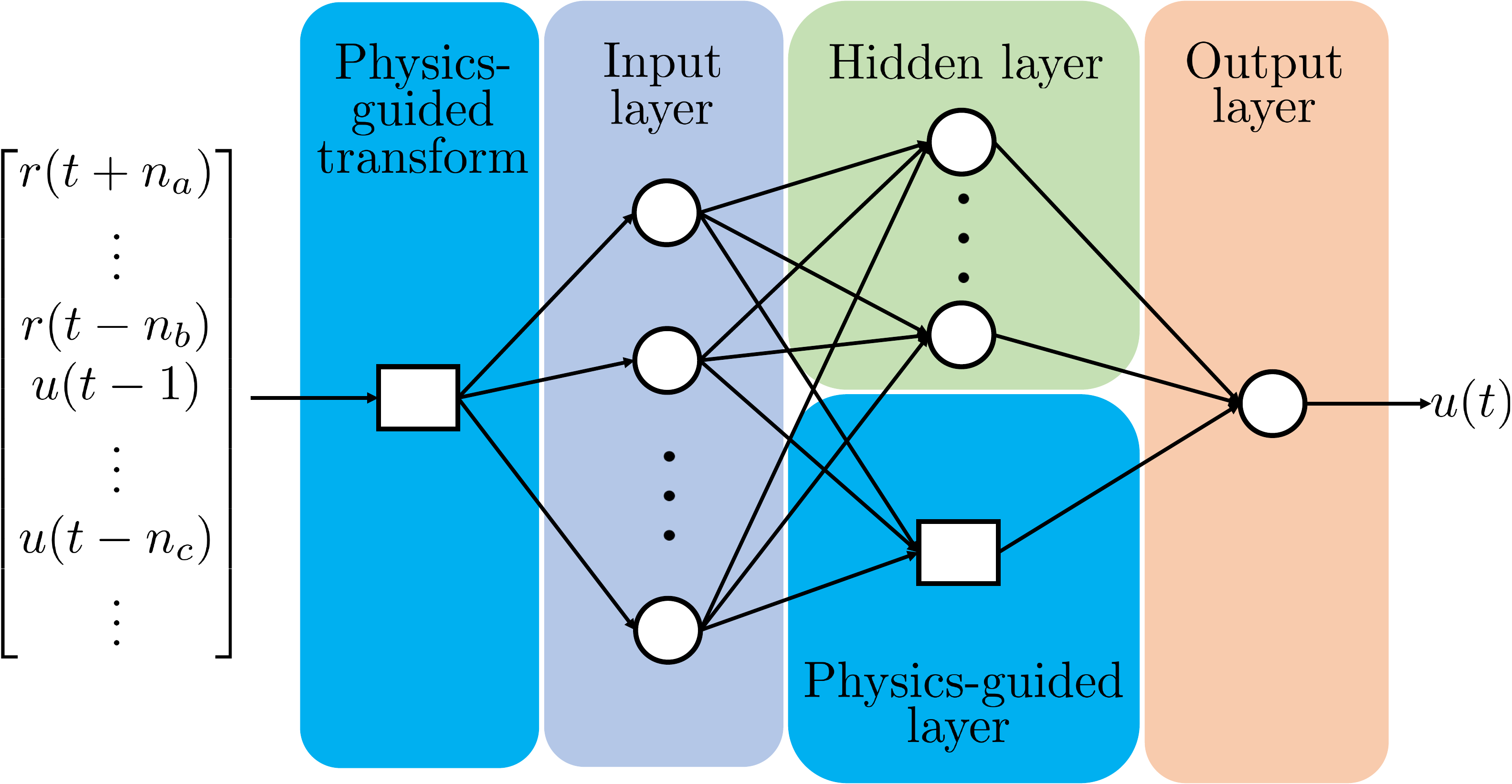}
	\caption{Illustration of a Physics--Guided Neural Network architecture.}
	\label{fig:PGNN}
\end{figure}

The PGNN--I is obtained by using the black--box NN structure, see Fig.~\ref{fig:NNARX}, in combination with a physics--guided input transformation, indicated in Fig.~\ref{fig:PGNN}. The PGNN--I and NNARX are both presented with the same information, but the physics--guided input transformation provides the PGNN--I with data that is better suited to approximate the considered system, and thereby guides the training. 

Let us consider the general CLM feedforward controller \eqref{eq:CLMFeedforward}, from which we can see that $u(t)$ is computed using a combination of the previous input $q^{-1} u(t)$, ZOH discretized acceleration $\frac{2}{T_s^2}q(1-q^{-1})^2 r(t)$, and current and past friction $(1-q^{-1}) F_{\textup{fric}}(r(t), \dot{r}(t))$. 
The approach in \cite{Karpatne2017}, inserting a parametric model computed output in the NN input, cannot be used, since we assume the mass and friction to be unknown. 
However, we do know that the friction depends on position, velocity and its sign. For this reason, we choose the PGNN--I feedforward controller
\begin{align}
\begin{split}
    \label{eq:PGNNIFeedforward}
    u(t) =&  f_{\textup{PGNN--I}} \Big(\ddot{r}(t), u(t-1), r(t), r(t-1), \dot{r}(t), \hdots \\
    & \hdots, \dot{r}(t-1), \textup{sign} (\dot{r}(t)), \textup{sign}(\dot{r}(t-1)) \Big).
\end{split}
\end{align} 
\textcolor{black}{Similar to $\dot{r}(t)$, above $\ddot{r}(t)$ stands for $\frac{2}{T_s^2} q (1 - q^{-1})^2 r(t)$, i.e., a second order discretization operator corresponding to ZOH. }

The physics--guided input transformation is similar to the use of the non--causal basis functions in feedforward control tuning, as proposed in \cite{Blanken2020b}. The main difference with respect to the developed PGNN--I lies in the additional hidden layer, which provides the means to approximate forces not captured by the physics--guided input transformation, such as the Stribeck friction and force ripple. Further unknown parasitic forces are expected to be present in the experimental setup. 

\subsection{PGNN--II: Physics--Guided Structure Design}
The PGNN--I uses physical knowledge to transform the input, simplifying the approximation of the system under consideration. 
Physical knowledge however is often available in the form of a specific structure of the system, e.g., a linear effect of friction components or parameter ranges, e.g., positive friction coefficients. 
For this reason, a physics--guided layer is introduced parallel to the hidden layer, see Fig.~\ref{fig:PGNN}, and the resulting physics--guided NN is referred to as PGNN--II. 

\textcolor{black}{The output of the physics--guided layer is defined as $x_{\textup{PGL}} = T(x_1)$, with $T(\cdot)$ any nonlinear transformation to embed system dynamics. Additionally, we recall that $x_i$ denotes the output of layer $i$ and define $x_2 = f_a (x_1)$, which is obtained by using \eqref{eq:OutputMLP} with $i=2$. Then, the output layer of the PGNN--II yields}
\begin{align}
    \begin{split}
        \label{eq:PGNNIIFeedforward}
        u(t) = W_{3,2} f_a(x_1) + W_{3,\textup{PGL}} T(x_1) + B_3 =:f_{\textup{PGNN-II}}(x_1),
    \end{split}
\end{align}
\textcolor{black}{with $x_1$ the physics--guided transformed input, see Fig. \ref{fig:PGNN}. Choosing the same physics--guided input transform as for the PGNN--I in combination with \eqref{eq:PGNNIIFeedforward} gives the  PGNN--II feedforward}
\begin{align}
\begin{split}
    \label{eq:PGNNIIFeedforward2}
    u(t) =&  f_{\textup{PGNN--II}} \Big(\ddot{r}(t), u(t-1), r(t), r(t-1), \dot{r}(t), \hdots \\
    & \hdots, \dot{r}(t-1), \textup{sign} (\dot{r}(t)), \textup{sign}(\dot{r}(t-1)) \Big).
\end{split}
\end{align}

\begin{remark}
\textcolor{black}{ The structure of the friction compensator \eqref{eq:FrictionCompensatorFeedforward} is embedded in the PGNN--II structure defined in \eqref{eq:PGNNIIFeedforward2}. This can be seen by using a direct feedthrough for the physics--guided layer, i.e., $T(x_1) = x_1$ in \eqref{eq:PGNNIIFeedforward}, choosing hidden layer weights zero, i.e., $W_{3,2} = 0$, and physics--guided layer weights $W_{3,\textup{PGL}} = \begin{bmatrix} 0 & \hat{m} & \hat{f}_v & \hat{f}_v & \hat{f}_c & \hat{f}_c \end{bmatrix}$. Hence, we expect the PGNN--II structure to increase performance and enhance extrapolation capabilities, which is confirmed by the simulation results in Sec.~\ref{sec:4}. }
\end{remark}

\textcolor{black}{The above remark shows that the physics--guided layer can be chosen such that its corresponding weights have a physical interpretation. Therefore, constraints can be imposed on the physics--guided layer weights, e.g., positive mass $\hat{m} > 0$. Also, training convergence can be improved by using parameter knowledge to initialize the physics--guided layer weights.} 
For assessing the performance improvent of PGNN--II with respect to the friction compensator \eqref{eq:FrictionCompensatorFeedforward}, in what follows we implement \eqref{eq:PGNNIIFeedforward} with $T(x_1) = x_1$  and we use the initial values $W_{3,\textup{PGL}} = \begin{bmatrix} 0 & \hat{m} & \hat{f}_v & \hat{f}_v & \hat{f}_c & \hat{f}_c \end{bmatrix}$ for the physics--guided layer weights. Then the resulting PGNN--II is trained using standard algorithms as in Sec.~\ref{sec:Benchmarks}, resulting in optimal weights for the full NN.


\section{Performance Evaluation and Comparison with Benchmark Feedforward Controllers}
\label{sec:4}
Data generation and performance evaluation are done based on a realistic Matlab Simulink model of the CLM, which includes a sinusoidal commutation algorithm and an identified Fourier model of the electromagnetics (see \cite{Nguyen2018} for more information); hence, the realistic model will exhibit a ripple force. 

The NNARX, PGNN--I and PGNN--II are trained with the MATLAB NN toolbox, according to the training procedure described in Sec.~\ref{sec:Benchmarks}. The risk of getting stuck in a local minima is reduced by performing $M=50$ consecutive trainings using random weight initialization. 
Results are presented for $n=2$ and $n=4$ hidden layer neurons, because choosing $n>4$ does not improve performance, due to over parameterization. Activation functions are chosen based on their resulting tracking performance for the different controllers.  

The PGNN--I and PGNN--II are compared to the benchmark feedforward strategies in terms of:
\begin{enumerate}
	\item Training performance: value of the MSE on the test-data after training, i.e., the value of the cost function \eqref{eq:CostFunction} evaluated on the test data. 
	\item Tracking performance: the mean average error (MAE) $\frac{1}{N} \sum^N \mid e(t) \mid$ of the tracking error. 
\end{enumerate}

\begin{table}
	\caption{Evaluation of the MSE cost function on the test data--set.}
	\label{tab:TrainingConvergence}
	\centering
	\includegraphics[width=0.8\linewidth]{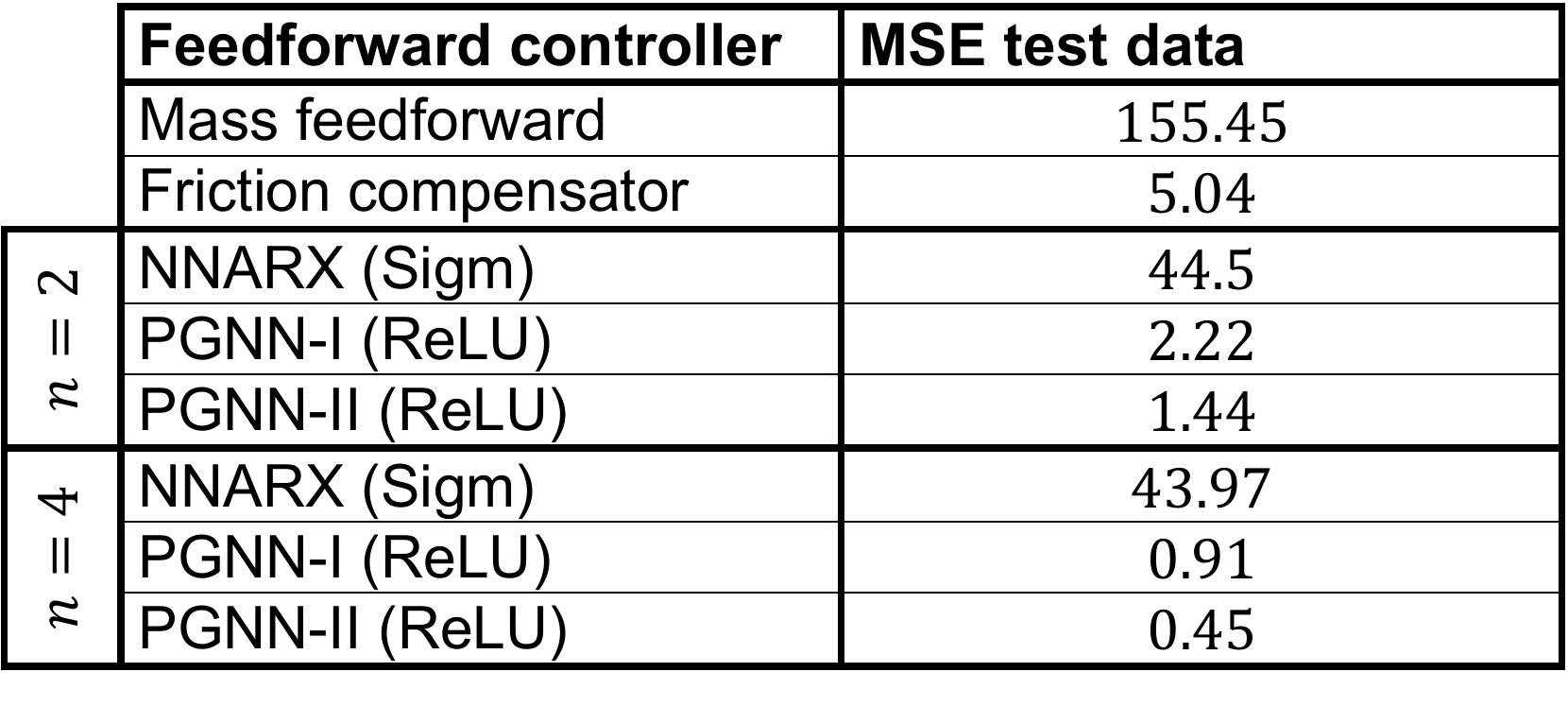}
\end{table}
\begin{table*}[ht]
	\caption{MAE of the tracking error for different strategies.}
	\label{tab:TestingPerformance}
	\centering
	\includegraphics[width=0.7\linewidth]{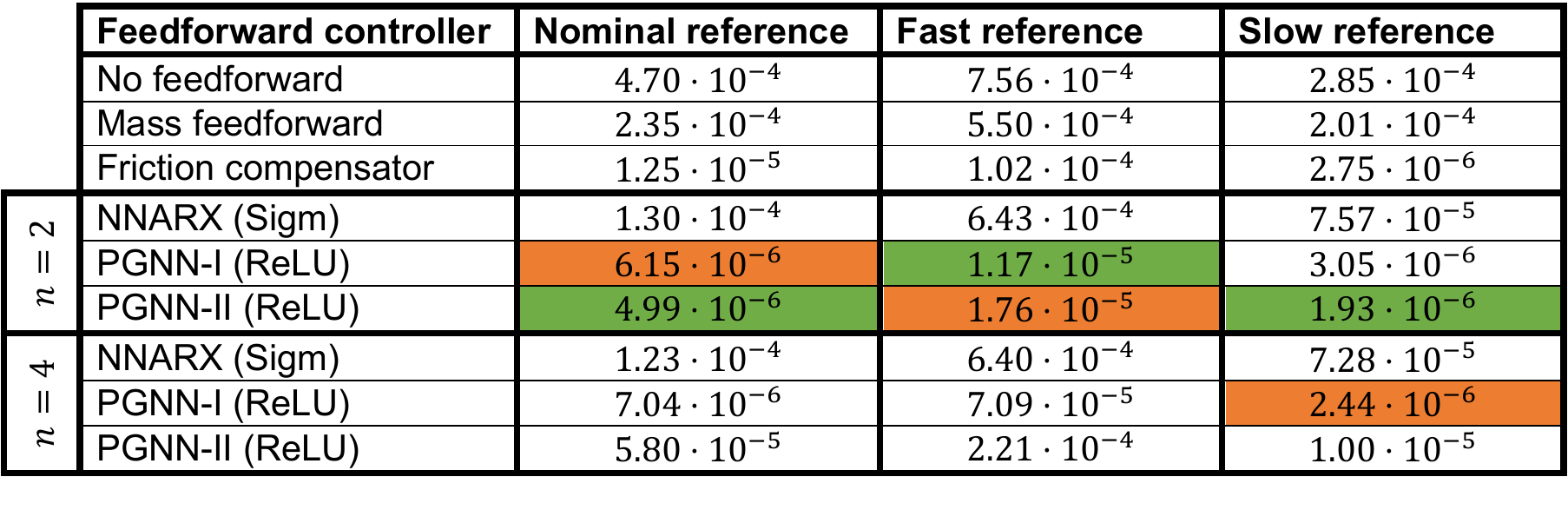}
\end{table*}

The training performance is summarized in Table~\ref{tab:TrainingConvergence}. This basically states the amount of input force that is not compensated for by the respective feedforward controller. For comparison, no feedforward gives an MSE of $8.22 \cdot 10^3$~$N^2$. 
From the mass--acceleration feedforward, we can see that approximately $98$\% of the required input force in the data set can be explained by inertial forces. From the remaining input, approximately $97$\% is assigned to the viscous and Coulomb friction. The NNARX fails to capture the general dynamics, as it is barely able to better reproduce the test data than the mass--acceleration feedforward. The PGNNs seem to reproduce the system dynamics best and obtain significantly better training performance than the friction compensator. 

The MAE is used to compare tracking performance of the feedforward controllers, on three reference trajectories:
\begin{enumerate}
    \item Nominal trajectory introduced in Sec. \ref{sec:TrainingData}:  $r = \left\{0, 0.05 \right\}$ $m$, $v_d = 0.05$ $\frac{m}{s}$, $a_d = 4$ $\frac{m}{s^2}$, $j_d = 1000$ $\frac{m}{s^3}$.
    \item Fast trajectory: $v_d = 4 \cdot 0.05$ $\frac{m}{s}$, $a_d = 4 \cdot 4$ $\frac{m}{s^2}$.
    \item Slow trajectory: $v_d = \frac{1}{4} 0.05$ $\frac{m}{s}$, $a_d = \frac{1}{4} 4$ $\frac{m}{s^2}$.
\end{enumerate}
The fast and slow trajectory are used to examine extrapolation capabilities of the feedforward controllers. 
The MAE values resulting from a single back--and--forth motion are summarized in Table \ref{tab:TestingPerformance}. The best and second best results are indicated with green and orange, respectively, for each trajectory.

Both the mass--acceleration feedforward and the friction compensator improve tracking performance of the CLM; the additional compensation of viscous and Coulomb friction makes the friction compensator outperform the mass--acceleration feedforward. 
The NNARX improves over the situation without feedforward, but is significantly outperformed by the friction compensator. 

Although training performance for the PGNN--I and PGNN--II was better for $n=4$, see Table~\ref{tab:TrainingConvergence}, $n=2$ yields better tracking performance. A possible reason is overfitting on the training data, which is, in contrast to the test references, dominated by inertial forces. Both PGNNs improve over the friction compensator on all references. 
The ReLU activation functions also allow the PGNN--I to extrapolate linearly to data unseen during training, such that performance is retained. Both PGNNs achieve the goal of having a MAE below $10$ $\mu m$ on the nominal trajectory. The key of improving tracking performance, is therefore within the addition of physical knowledge, rather than increasing the NN size. 

\begin{figure}
	\centering
	\includegraphics[width=0.9\linewidth]{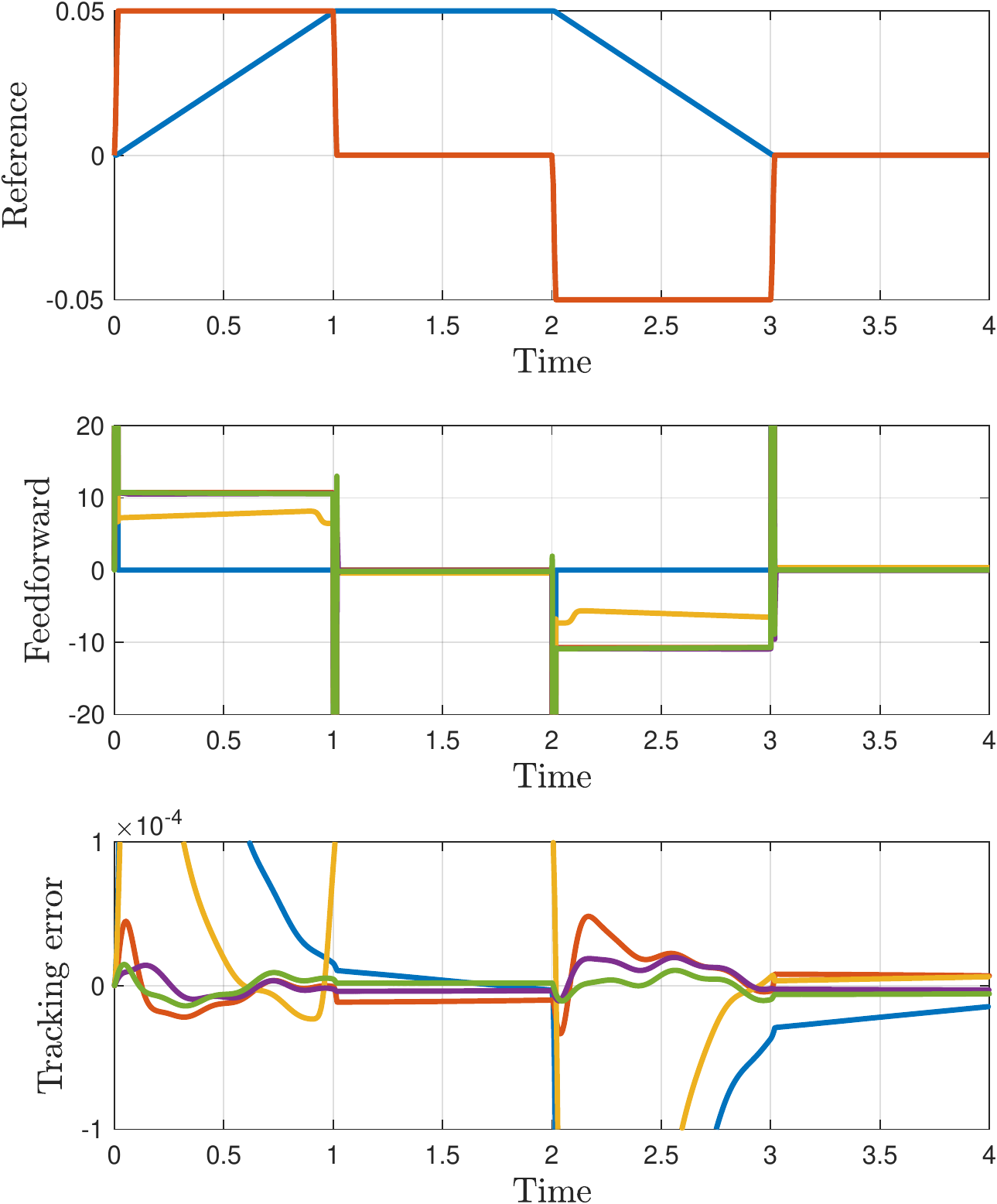}
	\caption{Results: mass--acceleration feedforward (blue), friction compensator (red), NNARX (yellow), PGNN--I (purple), and PGNN--II (green).}
	\label{fig:NNsFFGood}
\end{figure}

The different feedforward signals for the nominal reference trajectory are shown in Fig. \ref{fig:NNsFFGood} for the best performing feedforward controllers of each type. 
The NNARX generates a different feedforward signal during constant velocity, which might be related to the training converging to a local minimum. 
It is worth to mention that increasing the NNARX size did not yield better performance. Additionally, the close resemblance between the friction compensator, PGNN--I, and PGNN--II indicates the largely linear behaviour of the CLM when considering the nonlinear input transformation $\textup{sign}(\dot{r})$.
The friction compensator experiences an increase in tracking error at the beginning of a motion, as well as a slight offset when the velocity returns to zero. This indicates the inability to compensate for the Stribeck friction, that is only excited at small velocities. In contrary, the PGNNs seem to better capture this phenomenon, as they do not suffer from this offset. 

\section{Conclusions}
\label{sec:Conclusions}
In this paper, we investigated the use of NNs in inversion--based feedforward controller design for an industrial linear motor. A black--box NNARX was shown to fall short, as it was unable to capture general system dynamics and therefore did not improve tracking performance. In order to still exploit the universal approximation capabilities of NNs, this paper proposed the use of physical insights, obtained by a specific design of the NN inputs and structure. Compared to several benchmark feedforward strategies, a significant increase in performance was achieved. The limiting factor of tracking performance was shown to be the amount of physical knowledge embedded within the NN, rather than its size. As a next step, the developed PGNNs will be validated in real--time experiments on the available CLM setup.






\bibliographystyle{IEEEtran}
\bibliography{References}
\end{document}